\documentstyle[eqsecnum,preprint,tighten,aps]{revtex}
\newfont{\largemi}{cmmi10}
\newfont{\smallmi}{cmmi6}

\begin{document}
\draft

\title{Towards understanding the probability of $0^+$ ground states 
 in even-even many-body systems }
\author{ Y. M. Zhao$^{a,c,d}$\footnote{on leave from department of Physics,
 Southeast University, Nanjing 210018 China},
  and A. Arima$^{b}$}

\vspace{0.2in}
 \address{ $^a$ Cyclotron Center,  Institute of Physical Chemical Research (RIKEN), \\   
Hirosawa 2-1,Wako-shi,  Saitama 351-0198,  Japan\\
$^b$ The House of Councilors, 2-1-1 Nagatacho, 
Chiyodaku, Tokyo 100-8962, Japan \\
$^c$ Department of Physics, Saitama University, Urawa-shi, Saitama 338 Japan \\
$^d$ Bartol Research Institute, University of Delaware,  
Newark, DE19716-4793 USA  }

\date{\today}
\maketitle

\begin{abstract}
For   single-$j$ shells with $j=\frac{7}{2}, \frac{9}{2}$ and $\frac{11}{2}$, 
we relate the large probability of $I^+$ ground states
to   the largest (smallest) coefficients $\alpha^J_{I(v \beta)}$
$=$ $\langle nv \beta I |$ $A^{J \dagger} \cdot A^J | n v\beta I
\rangle$, where $n$ is the particle number, $v$ is the seniority,
$\beta$ is an additional quantum number, and 
$I$ is the angular momentum of the state.  
Interesting regularities of
the probabilities of  $I^+$ ground states are noticed and 
discussed  for 4-particle systems.
Several counter examples of the
$0^+$ ground state (0GS) predominance 
 are noticed for the first time. 

\end{abstract}


\vspace{0.4in}
        
\newpage

The  shell structure  and pairing  are ``universals" 
which play important roles   in  many-body systems
such as metal clusters, atoms, and nuclei etc.
Therefore, a  study which relates  intrinsic 
properties of the shell model  with the origin of the pairing  
phenomenon in many-body systems is extremely important. 
Recent work by Johnson
et al \cite{Johnson1,Johnson2,Johnson3} was the first effort
along these lines. In that work, 
 the low-lying spectra of many-body systems with even
 number of particles  were examined
using  random two-body interactions (TBRE),
and its results showed a preponderance of $I^{\pi}=0^+$ ground states
(0GS). This phenomenon also occurs  in  $sd$-boson systems, 
and was referred to as ``a robust property" of the
valence space  \cite{Bijker,Casten}. All these studies  
demonstrate that the features of pairing 
arise from a very large ensemble of two-body  
interactions and seem to be independent 
 of  the specific character of the force. 
These  observations, however, seem to be  contrary  
to the traditional assumption in nuclear physics,  whereby    
the 0GS dominance in even-even nuclei is 
a reflection of a strong pairing interaction associated with 
a strong short-range attraction between identical nucleons.  

It is therefore very important  and interesting to study
why the  0GS are favored 
in  even-even nuclei when using the TBRE
 \cite{Johnson1,Johnson2,Bijker1}.
There have been several efforts to 
 understand this phenomenon.  
In Ref. \cite{Bijker1}  an analysis was carried out on  
 systems of identical particles occupying orbits with 
$j=\frac{1}{2}, \frac{3}{2},$ and $\frac{5}{2}$, and it was 
indicated that there is a correspondence between a larger
distribution width  of $0^+$  states  and the
  0GS dominance.  In another paper, 
it was  proposed \cite{Bijker2} that for a
system of interacting bosons the probability that 
the ground state has a certain value of the angular momentum 
is not really fixed by the full  
 distribution  of eigenvalues, but rather by that of the lowest one. 
In \cite{zelevinsky}, Muhall  et al. discussed the 0GS dominance 
 within single-$j$ shells by using geometric chaoticity and
 uniformly changed random  interactions.   In \cite{kus}, Kusnesov
 discussed the $sp$-boson case. 
In \cite{Johnsonx} Kaplan et al. 
 studied the correlation between the 
eigenvalues and spins of the states in some simple cases. 

All these studies [1-10] on the   0GS dominance  are 
interesting and important, and have 
potentially impacted our understanding on
the origin of one of the most  
characteristic features of nuclear spectra. 
An essential  understanding of the   0GS  predominance, however,
 has not yet been  achieved. 
 
Towards that goal, we present in this Rapid Communication 
a new view on  the origin of the   0GS dominance.
We are able to achieve this new understanding because
we focus on the simplest non-trivial problem, that  of
identical nucleons in a single-$j$ shell. This problem
has the advantage that  it is simply solvable, thus permitting
useful insight to be obtained which can hopefully
then be applied to more general
problems.
  In the process, some previously unrecognized 
 features, such as counter examples of the   0GS 
predominance, and regularities of probability of $I^+$ 
ground states, have been discerned. 

We first discuss the properties of linear combinations of
random numbers. Let $G_J$ be a set of 
random numbers with a distribution function 
\begin{equation}
  \rho(G_J) = \frac{1}{\sqrt{\pi}} exp(-G_J^2),~~~J=0,2, \cdots 2(j-1),  \label{tbre}
\end{equation}
and $F(k)$ be a set of linear combinations 
of $G_J$:
\begin{equation}
F(k) = \sum_J \alpha_k^J G_J, ~~~ k = 1, 2, \cdots N,    \label{fk}
\end{equation}
where $j$ labels the single-$j$ shell, and
 $N$ is the total number of states (to be introduced later).  
It can be shown  that distribution functions  of random $F(k)$ are  
\begin{equation}
  \rho(F(k)) = \frac{1}{\sqrt{\pi}} exp(- \left( F(k) \right)^2/g_k^2),
 ~~ g_k^2 = \sum_J \left(\alpha_k^J\right)^2. 
   \label{randomx}
\end{equation}
If $\alpha_m^{J'}$ in Eq.~(\ref{fk})
is the largest (or the smallest) among all the 
$\alpha_k^{J'}$ ($k =1, \cdots   N$), the  probability of 
 $F(m)$ being both the smallest and the largest number  is large. 
 To show this, let us look at  
\begin{equation} 
 {\cal F}(k) =  F(k) - F(m)  = 
 c_k^{J'}   G_{J'} + \left( \sum_{J \neq J'} 
\left( \alpha_k^{J} - \alpha_m^{J}  \right) G_J \right)
 =  c_k^{J'}   G_{J'} +  {\cal F}'(k),   \label{shift}
\end{equation}
where $k \neq m,$ 
 and  $c_k^{J'}= \left( \alpha_k^{J'}  - \alpha_m^{J'} \right)$,
 ${\cal F}'(k) = \sum_{J \neq J'} \left( \alpha_k^{J}
 - \alpha_m^{J}  \right) G_J$. The right hand side of  Eq.~(\ref{shift}) has two
 terms, both of which are Gaussian type random numbers. The coefficient 
 $ c_k^{J'}$ is  negative or positive, and thus effectively produces
 a shift in $ {\cal F}(k)$, as is evident from Eq.~(\ref{shift}). 
Therefore,  all of the functions ${\cal F}(k)$   
have large probabilities to be both negative and positive,  
depending on the sign of the shift, i.e., 
 $F(m)$ has a large probability to be both the
smallest and the largest. 

Roughly speaking, one may   use  
  $d=|c_{k'}^{J'}|$, 
 which is the   smallest among all the
$|c_{k}^{J'}| (k \neq m)$, as an estimate of 
the shifts, and  use the maximum of 
$ D= \sqrt{\sum_{J \neq J'} 
\left( \alpha_k^{J} - \alpha_m^{J}  \right)^2}$ 
 as an estimate of 
the distribution widths of the functions $ {\cal F}'(k) $. 
The probability of $F(m)$ being the smallest
(largest) is very large when $d$ is comparably large ($d\sim D$).  
This  occurs when there are a very few  (3$\sim$6)  parameters in the Hamiltonian.  
 The probability  of $F(m)$ being  the smallest (largest)  
 is  roughly determined by $g_k$,  the 
 widths of the quantities $F(k)$, 
 when $d << D$. The states with large eigenvalue widths   
 have large probabilities  to be both the smallest and 
 the largest. This situation occurs if  there are many (e.g., more than 20)
   TBRE parameters. In Ref.  \cite{Bijker1}, the 0GS probability are determined
   by the total widths of the energy eigenvalues because there are
   30 parametrized T=1 interactions. 

  If there are two or more   
  coefficients $\alpha_m^{J'}$ ($J'=0,2,\cdots 2j-1$)
  which are the largest or smallest  for different functions $F(k)$, 
   the probability of  finding  $F(m)$ as both  
   the smallest and the largest increases. 

Keeping the above ideas in mind, 
we now return to the shell model calculation  for single$-j$ shells. 
The Hamiltonian for a single$-j$ shell  is
\begin{eqnarray}
&& H = \sum_J \sqrt{2J+1} G_J
\left[ A^{J \dagger} \times
\tilde{A}^J \right]^0, \nonumber \\
&&  A^{J \dagger} = \frac{1}{\sqrt{2}} \left[ a_j^{\dagger} \times a_j^{\dagger}
 \right]^J,
 \tilde{A}^J = - \frac{1}{\sqrt{2}} \left[ \tilde{a}_j \times
 \tilde{a}_j \right]^J, G_J = \langle j^2 J|V| j^2 J \rangle.  \label{pair}
  \nonumber
\end{eqnarray}
The parameters $G_J$  are distributed according to   Eq.~(\ref{tbre}). 

Consider the simplest  
example, a $j=7/2$ shell with 4 particles, where  
 all  the states are labeled by  
 their total angular momenta $I$ and their seniority  
quantum numbers ($v$). The eigenvalues $E_{I(v)}$ are as follows \cite{Lawson}:  
\begin{eqnarray}
&&E_{0(0)} = {\bf \frac{3}{2}} G_0 +  \frac{5}{6} G_2 +
              \frac{3}{2} G_4 + \frac{13}{6} G_6, 
E_{2(2)} =\frac{1}{2} G_0 +  \frac{11}{6} G_2 +
              \frac{3}{2} G_4 + \frac{13}{6} G_6, \nonumber \\
&&E_{2(4)} = ~~ ~~~~~~            G_2 +
              {\bf\frac{42}{11}} G_4 + {\it \frac{13}{11}} G_6, 
E_{4(2)} =\frac{1}{2} G_0 +  \frac{5}{6} G_2 +
    ~          \frac{5}{2} G_4 + \frac{13}{6} G_6, \nonumber \\
&&E_{4(4)} = ~~~~~~         {\bf \frac{7}{3}} G_2 +
                  ~ {\it 1} \cdot   G_4 + \frac{8}{3} G_6, 
E_{5(4)} =  ~~~~~~~            \frac{8}{7} G_2 +
              \frac{192}{77} G_4 + \frac{26}{11} G_6, \nonumber \\
&&E_{6(2)} =\frac{1}{2} G_0 +  \frac{5}{6} G_2 +
     ~         \frac{3}{2} G_4 + \frac{19}{6} G_6, 
E_{8(4)} = ~~    {\it \frac{10}{21}} G_2 +
              \frac{129}{77} G_4 +  {\bf \frac{127}{33}} G_6,  \label{7/2}
\end{eqnarray}
where  bold font is used for the largest and italic
for the smallest amplitudes in an expansion in terms of $G_J$,  
\begin{equation}
E_{I(v)} = \sum_J \alpha^J_{I(v)} G_J   \label{alpha}. 
\end{equation}
Using  Eq.~(\ref{7/2}), 
we can predict   the probability of $I^+$ ground states 
  without carrying out energy calculations with the TBRE. 
  For example,  the  0GS  probability
  is determined by 
\begin{eqnarray}
&& \int dG_0 \int dG_2  \int dG_4
\int dG_{6}
  \int dE_{0(0)}   
\int_{E_{0(0)}} dE_{2(2)}
\cdots  \int_{E_{0(0)}} dE_{8(4)} 
\nonumber   \\ 
&&  \delta \left(E_{0(0)} - \sum_J \alpha_{0(0)}^J G_J \right) \cdots
  \delta \left(E_{8(4)} -   \sum_J \alpha_{8(4)}^J  G_J\right)
 \rho(G_0)  \rho(G_2)  \rho(G_4) \rho(G_6). \label{exact}
\end{eqnarray}
The probabilities  of finding $I$ as the spin of
 ground states  for a
 4-particle system with $j=\frac{7}{2}$
  are listed in Table I. The row  ``test"  corresponds to
results obtained by using Eq.~(\ref{7/2}) and 1000 
sets of the TBRE. The  
row ``cal." corresponds to the probabilities predicted  
by an  integral for each $I^+$ state similar to Eq.~(\ref{exact}) for the
 $0^+$ state.
The probabilities  calculated using the TBRE  and those predicted 
 by using  integrals like Eq.~(\ref{exact})  are all consistent with one another. 

 Using Eq.~(\ref{7/2}) and Eq.~(\ref{randomx}),
 $g_{I(v)}-$the distribution  width of $E_{I(v)}-$is obtained and  
  listed in the last row of  Table I. 
  It is found that  there is no 
 correspondence  between  the width $g_{I(v)}$  
  and the probability of finding $I^+(v)$ to be the ground state. 

Because there are only 4 parameters  in Eq.~(\ref{7/2}),
we can use the shift argument     
to predict which states have large probabilities to be the
ground.  They are the states with the largest (smallest) 
$\alpha_{I(v)}^J: I(v)$=0(0), 2(4),
4(4), 8(4). 

Our success in explaining which states
   can have  large
probabilities to be the ground state 
 for the $j=\frac{7}{2}$ shell
 encouraged us to go to higher $j$ shells.
The study of 4 particles in the $j=\frac{9}{2}$ shell 
is very interesting because 
the 0GS  is predominant in this case (See Fig. 1). 
This is therefore a good case in which to check  whether
the above assumptions continue to apply. 

Unfortunately, the eigenstates of 4 particles in this shell 
are complicated 
by mixings between states with  the same and 
 different  seniorities. In such a case  there  is no simple relation
between the eigenvalues and the two-body interaction 
parameters, $G_J$.    

To simplify the analysis for the $j=\frac{9}{2}$  case, we assume
that the  mixings
are negligible, and $\langle n v \beta I|  
 \sum_J \sqrt{2J+1}\left[ A^{\dagger J} \times
\tilde{A}^J \right]^0 | n v \beta I \rangle$ are 
the "eigenvalues".  This assumption is supported by the fact that 
in the basis  $| n v \beta I \rangle$ \cite{Zhao}  mixings are observed 
to be  small for $I>0$ states in even-$N$ systems in
 the $j=\frac{9}{2}$ shell.
It is also supported by a recent claim \cite{Johnson2}  that a 
 many-body system  picks out seniority as an approximate  
quantum number, even though it is not obviously implicit in the 
Hamiltonian.  If we could take those mixings into account 
in a simple way, those  off-diagonal matrix elements 
would presumably pull down the lower "eigenstates" even lower, and push up the 
higher ones even higher.  For $I=0$ states, 
the off-diagonal matrix elements are  comparable in magnitude to 
the diagonal matrix elements, and the 
pulling-down  and pushing-up effects should  be more important than
for $I \neq 0$ states.  Therefore, this assumption leads to an 
underestimate of the 0GS probability.

The coefficients $\alpha_I^J$  for each $G_J$ in the ``eigenvalues" 
are listed in Table II. There are 
three largest  $\alpha_I^J$ ($J=0, 4, 6$)  and one smallest $\alpha_I^J$
($J=8$)in the $I=0$ states. 
 Therefore, the 0GS  probability  must be very large. 
In addition, there are one largest $\alpha_I^J$ ($J=2$) and one smallest
$\alpha_I^J$ ($J=6$)  in one of the $4^+$ states. Finally, there
are one largest $\alpha_I^J$ ($J=8$)
and two smallest $\alpha_I^J$ ($J=2, 4$) in the
highest angular momentum state ($I=12$). 
This indicates large  probabilities  of $4^+$ ground states 
and $12^+$ ground states.  In one of the $2^+$ states   
($v=2$) the  coefficient of $G_6$ is very near to the smallest.
Thus, there should be a sizable  percentage of  $2^+$   ground states as well. 

Shell model calculations based on the TBRE 
 are   consistent with  the above discussion: 
The  probability of $0^+$ states to be the ground state
is  66.4$\%$, that of  $4^+$ is
12.2$\%$ and that of  $12^+$ is 17.4$\%$. 
The probability of finding $2^+$ states as the ground
state is  $3.4\%$. 
There are very small probabilities ($0.6\%$ in total)
for all other $I^+$ ground states. The predicted 0GS 
probability obtained   using an integral similar to Eq.~(\ref{exact}), 
where  no mixing is assumed   (which is exact for $j=\frac{7}{2}$),  
is  45.11$\%$.
As expected, this is smaller than the probability ($66.4\%$) found
when the mixings are taken into account. 
We note without further details that
the above argument is also  applicable  to 
$4$  and $6$ particles  in the  $j=\frac{11}{2}$ shell, and to the
general features of  odd-A fermion system  in 
small single-$j$ shells and $sd$-boson systems \cite{Zhao}.

We have also carried out several other sets of calculations, both for larger single-$j$
shells and in some cases for two-$j$ shells. Some of the
principal outcomes are:

1) The probability of  ground states with odd angular momenta 
 is  much smaller than that of their neighboring  even
 angular momenta even if the number of states
 with the same angular momentum $I$ found in $j^n$ configuration
  is   comparably large.

2) The unique $I_{max}$ state has a large probability  
to be the ground state, although 
 this probability  decreases  with $j$.
  According to  ref.  \cite{zelevinsky}, 
 by using random interactions which  distribute  
 uniformly   between $-1$ and 1, the 
  probability of $I^+_{max}$ ground states 
   staggers rapidly and  becomes  0 for several
   single $j$.

3) The $I=2^+$ and $4^+$ states have  large probabilities
 to be  the ground state.   This  indicates   
that  small and even angular
momentum  states are favored as the ground state in
single-$j$ shells.

4) The 0GS predominance obtained using the TBRE is not a ``rule"
without exceptions. 
There are counter examples: 
$j=\frac{7}{2}$   and $j=\frac{13}{2}$ shells  with   
4 particles (refer to Fig. 1);
two-$j$ shell ($j = \frac{7}{2}$ plus $\frac{5}{2}$)
with  4 particles,   
where the probability of the 0GS is $24\%$ while that
of the $2^+$ GS is  $33\%$.  
                                 
A summary of the above regularities is  
shown in Fig. 1, from which we see that  
when the 0GS probability  is a maximum, the
probability of $2^+$ ground states is a minimum, 
and vice versa. 
Fig. 1  indicates that 
the  fluctuation of the 0GS probability
decreases with $j$, 
the 0GS probability  saturates around 35$\%$, and   
the GS probabilities of $I=2,4$ and $I_{max}$  states decrease and  
converge around $9\%$. 
These regularities  are worthy of further study. 

To summarize,   we have presented in this Rapid Communication an
interpretation of the 0GS probability  
in terms of the largest and smallest coefficients 
 $\alpha_{I(v)}^J$   for several  single-$j$ shells. The interpretation
 applies when seniority is either a conserved or an approximately
 conserved quantum number.  Further study showed that
 this method  is not only applicable  to even number of fermions 
 in a small single-$j$ shell, but also is a  good benchmark
 to explain the $I$GS probability of odd-A fermions in a small single-$j$
 shell and general features of $sd$-boson systems \cite{Zhao}.
 Therefore, for the first time, the $I$GS  probabilities
 of a variety of systems
 can be discussed on the  same footing using the method proposed
 in this paper.

The 0GS probabilities  have been  calculated   using 
integrals like Eq.~(\ref{exact}),
    the predicted probabilities 
 being  reasonably consistent with those obtained   using
   the TBRE. 
We  showed that the width of the distribution of $0^+$ eigenvalues is not 
a manifestation of the 0GS preponderance in  small  single-$j$ shells.

Two other conclusions not yet mentioned  are as follows:
1) If  $I^+$ states  have a very large probability to be the ground state, 
 there must be, on  
  average, a  large  energy  gap
between the $I^+$ ground states and other states. This gap 
comes   from a large shift $d$ in  small $j$ shells, 
where there are  very few parameters,  or  from the  difference between 
the width of  eigenvalues for the $I^+$ states  and those for 
all other angular momentum  states  in  very large single-$j$   
shells and multi-$j$ shells.  
2) If  certain $I^+$ states are favored to be the  
ground state, they are  also favored to be the highest state with the
same probability.

As $j$ increases, the approximations made in this study, 
 i.e., the omission of  mixings between states with different 
seniorities and  those 
among  states with the  same seniority deteriorate.
In principle, the relationship between the
 eigenvalues and the two-body interactions is
{\bf not} linear.   As a   consequence, more general
or more complicated cases may not be studied in this simple way. Nevertheless, our
analysis should provide 
 helpful clues for what is
going on with regards to 0GS predominance in more general even-even many-body systems.

Finally, it is not yet understood why certain coefficients 
 $\alpha^J_{0(v)}$ turn out to be 
the largest and/or the smallest among all of them. 
Since this  observation seems to be critical to explaining 0GS dominance,
at least in small single-$j$ shells, further consideration of this issue is warranted.

  We are grateful to Dr. N.Yoshinaga for his 
 generosity  in allowing us to use his 
modified  ``two-$j$ shell"  program. 
Many discussions with  Drs.  N. Yoshinaga, 
S. Pittel, and S. Yamaji,  and
communications with Dr. C.W. Johnson 
are gratefully acknowledged. We thank Dr.
S. Pittel for his careful reading of the mansucript.
This work is supported in part by the Japan Society
for the Promotion of Science
(contract ID: P0102) and the U.S. NSF under grant No.
PHY-9970749.

\newpage

{\bf Caption}: 

Fig. 1. Probabilities of $I^+$ ground states for 
different $j$ shells with 4 particles.
All probabilities are obtained from 
1000 runs of the TBRE.

\newpage
\vspace{0.1in}

{TABLE I. The probability  of each state to be  the ground
state and its distribution width for 4 particles in a  $j=7/2$ shell. 
All eigenstates are uniquely 
labeled by their seniorities and angular momenta. 
The first line specifies the 
angular momentum and  seniority  for each state. 
The probabilities labeled by ``test" are obtained from  1000  runs  
 of the TBRE, and those  
 labeled by ``cal." are obtained by integrals  like 
 Eq.~(\ref{exact}) for $0^+$ state (see the text for further details). 
 The  distribution  width,   $g_{I(v)}$, of each energy level
 is listed in the 3rd row.  The shift, $d$, of each level is
 given in the last row. The  $d$ inside a bracket corresponds
 to a shift coming from the smallest $\alpha_{I(v)}^J$. }

\begin{tabular}{ccccccccc} \hline \hline
$I(v)$ & 0(0) & 2(2) & 2(4) & 4(2) & 4(4) & 5(4) & 6(2) & 8(4) \\  \hline
test 
& $19.9\%$ & $1.2\%$ & $31.7\%$ & $0.0\%$ & $25.0\%$ &
$0.0\%$ & 0.0$\%$  & 22.2$\%$ \\
cal. & 18.19$\%$ & 0.89$\%$ & 33.25$\%$ & 0.00$\%$ &
22.96$\%$ & 0.00$\%$ & 0.02$\%$ & 24.15$\%$ \\
$g_{I(v)}$ & 3.14 & 3.25& 4.12        & 3.45 & 3.68       & 3.62 & 3.64 &4.22  \\
$d$        & 1.00 &   - & 1.32 (0.98) &  -   & 0.50(0.50) & -    &  -   & 0.68(0.35) \\ \hline \hline
\end{tabular}

\vspace{0.3in}

{TABLE II. The coefficients $\alpha_{I(v\beta)}^J$ 
for 4 particles in a $j=\frac{9}{2}$ shell. 
 Bold font is used   for the largest
 $\alpha^J_I$ are the largest and italic for
 the smallest  $\alpha^J_I$.  }

\begin{tabular}{ccccccc} \hline  \hline
$I$ & v &  $G_0$ &  $G_2$ &  $G_4$ &  $G_6$ &  $G_8$  \\  \hline 
0& 0 &  {\bf 1.60} & 0.50 & 0.90 & 1.30 & 1.70 \\
0& 4 &  0.00 & 0.20 & {\bf 2.57} & {\bf 2.91} & {\it 0.32} \\ 
2 & 2 &  0.60 & 1.43 & 1.22 & 0.893 & 1.86 \\
2&4 &  0.00 & 1.35 & 1.69 & 1.70 & 1.26 \\  
3 & 4 &  0.00 & 0.36 & 2.28 & 2.63 & 0.71 \\ 
4& 4 &  0.00 & 0.50 & 2.08 & 2.43 & 0.99 \\
4& 4 & 0.00 & {\bf 2.04} & 1.02 & {\it 0.890} & 2.06 \\
4& 2 & 0.60 & 0.68 & 1.04 & 2.40 & 1.28 \\  
5& 4 & 0.00 & 1.00 & 1.59 & 1.84 & 1.57 \\ 
6& 4 & 0.00 & 1.64 & 0.98 & 1.08 & 2.29 \\
6& 4 & 0.00 & 0.39 & 1.85 & 2.34 & 1.43 \\
6& 2 & 0.60 & 0.34 & 1.66 & 1.33 & 2.07 \\  
7& 4 & 0.00 & 1.20 & 1.09 & 1.40 & 2.31 \\  
8& 4 & 0.00 & 0.41 & 1.42 & 2.05 & 2.13 \\
8& 2 & 0.60 & 0.55 & 0.68 & 1.58 & 2.59 \\  
9& 4 & 0.00 & 0.17 & 1.33 & 2.12 & 2.38 \\ 
10& 4 &  0.00 & 0.70 & 0.69 & 1.41 & 3.21 \\ 
12& 4 & 0.00 & {\it 0.00} & {\it 0.52} & 1.69 & {\bf 3.78} \\   \hline  \hline
\end{tabular}

\end{document}